\documentclass[preprint,emulateapj]{aastex}

\usepackage{ulem}

\slugcomment{June 2012, accepted for publication in ApJ}

\shorttitle{Glancing views of the Earth.}
\shortauthors{Garc\'ia Mu\~noz et al.}

\begin{document}

\title{Glancing views of the Earth.\\ 
From a lunar eclipse to an exoplanetary transit}

\author{A. Garc\'ia Mu\~noz\altaffilmark{1,2,4}, M. R. Zapatero Osorio\altaffilmark{3},}

\author{
 R. Barrena\altaffilmark{1,2}, 
P. Monta\~n\'es-Rodr\'iguez\altaffilmark{1,2}, E. L. Mart\'in\altaffilmark{3} \&
E. Pall\'e\altaffilmark{1,2}
}

\altaffiltext{1}{Instituto de Astrof\'isica de Canarias, C/V\'ia L\'actea s/n,
E-38205 La Laguna, Tenerife, Spain.}
\altaffiltext{2}{Departamento de Astrof\'isica, Facultad de F\'isica,
Universidad de La Laguna, La laguna, Tenerife, Spain.}
\altaffiltext{3}{Centro de Astrobiolog\'ia, CSIC-INTA, Ctra. de Torrej\'on a
Ajalvir, km 4, E-28550 Madrid, Spain}
\altaffiltext{4}{Email address: tonhingm@gmail.com}
 
\begin{abstract}

It has been posited that lunar eclipse observations may help predict
the in-transit signature of Earth-like extrasolar planets.
However, a comparative analysis of the two phenomena
addressing in detail the transport of stellar light through the 
planet's atmosphere has not yet been presented. 
Here, we proceed with the investigation of both phenomena by 
making use of a common formulation. 
Our starting point is a set of previously unpublished near-infrared
spectra collected at various phases during the August 2008 lunar eclipse.
We then take the formulation to the limit of an infinitely distant
observer in order to investigate the in-transit signature of the 
Earth-Sun system as being observed from outside our Solar System.
The refraction-bending of sunlight rays that pass 
through the Earth's atmosphere is a critical factor in the illumination of the
eclipsed Moon.
Likewise, refraction will have an impact on the 
in-transit transmission spectrum for specific planet-star systems
depending on the refractive properties of the planet's atmosphere, the stellar
size and the planet's orbital distance. 
For the Earth-Sun system, at mid-transit,  
refraction prevents the remote observer's access to the lower 
$\sim$12--14 km of the atmosphere and, thus, 
also to the bulk of the spectroscopically-active atmospheric gases. 
We demonstrate that the effective optical radius of the Earth in transit 
is modulated by refraction and varies by $\sim$12 km from mid-transit to
2nd contact.
The refractive nature of atmospheres, a property 
which is rarely accounted for in published investigations, 
will pose additional challenges to the 
characterization of Earth-like extrasolar planets.  
Refraction may have a lesser impact for Earth-like extrasolar
planets within the habitable zone of some M-type stars.

\end{abstract}

%\keywords{lunar eclipse --- transit --- Earth --- extrasolar planets ---refraction}
\keywords{}

\section{Introduction}

The contemplation of the Moon during a lunar eclipse reveals the dimming, 
and subsequent brightening, of the lunar disk as the satellite enters, and 
then exits, the shadow cast by the Sun-illuminated Earth. 
Ground-based observers have for decades constructed lightcurves of the 
sunlight reflected from the eclipsed Moon
\citep[e.g.,][and references therein]{link1962,garciamunozpalle2011}. 
The experiment is relevant to the investigation of the terrestrial
atmosphere because, for a known position of the Moon within the Earth's shadow,
the lightcurves are interpretable in terms of 
the optical properties of the atmosphere
\citep{ugolnikovmaslov2006,ugolnikovmaslov2008}. %intercepted by sunlight rays 
Beyond traditional photometric measurements,
two spectroscopic observations of the August 2008 lunar eclipse
have demonstrated the richness of the spectrum of sunlight reflected 
from the eclipsed Moon and shown that it contains the signature of the 
principal atmospheric constituents \citep{palleetal2009,vidalmadjaretal2010}.

Within the sample of known extrasolar planets, 
the transiting ones form a subgroup particularly apt for characterization. 
During transit, the planet partially blocks  the 
disk of its host star, causing a drop in the amount of stellar light
that arrives to the observer. 
The apparent stellar dimming 
is related to the sizes of the core and --when there is one-- 
the gaseous envelope of the planet.
The technique of in-transit transmission spectroscopy, which relies 
on comparing the apparent stellar dimming
at selected wavelengths, has led to the detection
of atoms and molecules such as Na, H, C, O, CO, CO$_2$, H$_2$O and CH$_4$ 
in the atmospheres of a few giant close-in extrasolar planets
\citep[e.g.][]{charbonneauetal2002,vidalmadjaretal2003, vidalmadjaretal2004,
tinettietal2007, desertetal2009,singetal2009,swainetal2009}. 
The same technique, applied to GJ 1214b, is helping elucidate 
whether this so-called super-Earth \citep{charbonneauetal2009} might contain abundant water in its
atmosphere \citep{bertaetal2012}.
 
Lunar eclipses and exoplanetary transits are related phenomena. 
In both instances a fraction of the light collected by the 
observer is stellar light
that has passed through a section of the planet's limb and, thus,
carries the signature of the planet's gaseous envelope.
The photon trajectories, whether direct or deflected in scattering collisions 
with the atmospheric constituents, 
determine the signature imprinted on the collected stellar light. 
Throughout the text, we use the term (lunar) eclipse to refer to the
alignment of the Sun, Earth and Moon, 
and transit to refer to the Earth's passage in front of the Sun as
observed from a remote vantage point.

The classical theory of lunar eclipses assumes that refracted sunlight,
rather than the scattered component, determines the brightness of the eclipsed Moon
\citep{link1962}.
However, \citet{garciamunozetal2011} have shown that 
both components may become comparably intense under conditions of elevated
aerosol loading.
With a few exceptions or brief references 
\citep{seagersasselov2000,brown2001,hubbardetal2001,huiseager2002, 
ehrenreichetal2006, kalteneggertraub2009,sidissari2010}, 
the effect of refraction on the lightcurves of transiting extrasolar planets has 
largely been ignored. The omission is likely due to the minor contribution of
refraction to the lightcurves of the close-in giant planets that constitute 
most of the extrasolar planets discovered to date.
However, and as shown below, refraction has a non-negligible impact for smaller 
planets far out from their host stars.

As the newly discovered extrasolar planets steadily approach 
terrestrial sizes
\citep[e.g.][]{charbonneauetal2009,legeretal2009,mayoretal2009,
batalhaetal2011,lissaueretal2011}, 
further attention is being devoted to the prediction of the 
in-transit signature of Earth-like extrasolar planets.
In preparation for future observations, 
a number of recent works have investigated the potential of 
in-transit transmission spectroscopy for the characterization of such targets
\citep[e.g.][]{ehrenreichetal2006,kalteneggertraub2009,
palleetal2011,raueretal2011}. 
In that context, it has been suggested that 
lunar eclipse spectra may help test 
the strategies that will eventually lead to the characterization
of Earth-like extrasolar planets \citep{palleetal2009, vidalmadjaretal2010}.

Our paper investigates the transport of stellar light through the
Earth's atmosphere in lunar eclipses and exoplanetary transits. 
Our main goal is to clarify the similarities and dissimilarities
in the signal collected by terrestrial observers in both types of event, 
emphasizing the impact of refraction.
The paper is structured in two main blocks. 
In the first block, {\S}\ref{theory_sec}--{\S}\ref{timevar_sec}, 
we introduce the lunar eclipse theory and 
investigate a set of previously unpublished 
near-infrared spectra of the August 2008 lunar eclipse. 
In the second block, {\S}\ref{transit_sec}, 
we take the lunar eclipse theory to
the limit of a remotely distant observer and address the comparison between 
lunar eclipses and transits. % over the 0.4--2.4 $\mu$m region of the spectrum. 
In {\S}\ref{timevar_sec}--{\S}\ref{transit_sec}, we also discuss the
conclusions drawn by some prior works in relation with the theory of lunar
eclipses and exoplanetary transits. 
Finally, {\S}\ref{summary_sec} summarizes the main conclusions. % of the work.
To the best of our knowledge, this is the first work that investigates the effect
of refraction on the in-transit spectral signature of Earth-like extrasolar 
planets.

%\newpage
\section{\label{theory_sec} The data. The lunar eclipse theory} 

The spectra investigated in {\S}\ref{timevar_sec} were obtained 
on 16 August 2008 
with the LIRIS spectrograph at
the William Herschel Telescope, WHT \citep{manchadoetal1998}, located at the 
Observatorio del Roque de los Muchachos (ORM),
La Palma, Spain. 
The observations 
were made alternating over two wavelength ranges with two different grisms: 
 $zj$, from 0.9 to 1.5 $\mu$m, and $hk$, from 1.4 to 2.4 $\mu$m. 
A total of 11 spectra of the Moon in umbra were collected, 
6 in $zj$ and 5 in $hk$. 
The umbra observations ran from 20:54 until 22:16UT 
and probed different phases of the Earth's inner shadow projected on the lunar
disk.
The data reduction and telluric correction follow \citet{palleetal2009} 
and will not be discussed further. 
The FITS raw datafiles are publicly available from the ING Archive.\footnote{http://casu.ast.cam.ac.uk/casuadc/archives/ingarch}
For their published spectrum from 1 to 2.4 $\mu$m, 
\citet{palleetal2009} merged the three spectra closest to greatest eclipse 
in each wavelength range. 
We will now focus on the individual umbra spectra in order to investigate
their evolution over time and distance from greatest eclipse.

For the interpretation of the observations, we produced model 
simulations of the eclipse spectrum. The simulations involved solving 
the radiative transport problem of
sunlight in the Earth's atmosphere according to 
the formulation laid out by \citet{garciamunozpalle2011}. 
The formulation assumes that the observer is located on the Moon's surface, 
which allows us to ignore the reflection of sunlight at the Moon 
and the atmospheric extinction above the Earth-bound observer. 

Generally,
we integrate the direct and scattered components of radiance
at the lunar observer's site 
over a solid angle $\partial \Omega$ that includes both the solar and planetary disks. 
Denoting by ${{L}}({\bf{x_O}}, {\bf{s_O}})$ the radiance at the observer's site 
$\mathbf{x_O}$ in the incident direction $\bf{s_O}$, the irradiance 
or flux of sunlight into an elementary surface oriented according to $\bf{n}_{\bf{x_O}}$
is obtained from:
\begin{equation}
F(\mathbf{x_O})= 
\int_{\partial \Omega}{{L}}({\bf{x_O}}, {\bf{s_O}}) \bf{s_O}\cdot n_{\bf{x_O}}
d\Omega(\bf{s_O}). 
\label{irradianceO_eq}
\end{equation}

For the direct component,  Beer-Lambert's law is integrated on the refraction-bent
trajectories of all possible lines of sight connecting 
the lunar observer's site with the solar disk. 
At the solar disk, the Sun's emission radiance is scaled according to a specified
limb-darkening function.
For the scattered component, the solar photon trajectories are simulated by a 
Monte Carlo algorithm. 
The formulation is general enough to investigate 
both the umbra and penumbra phases of a lunar eclipse and, as 
discussed below, to predict the in-transit signature of extrasolar planets. 

\begin{figure*}[h]
%\epsscale{1.80}
\noindent\includegraphics[width=40pc]{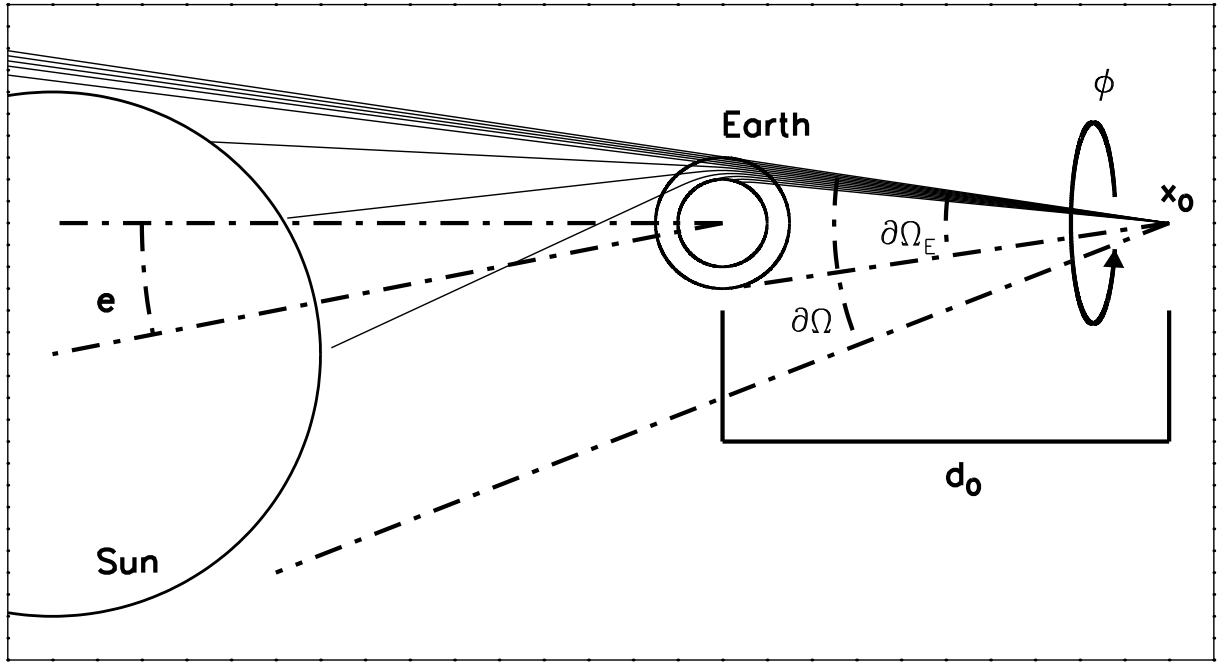}
\caption{\label{sketch0_fig} Sketch of the Earth-Sun system and the ray tracing scheme
from the observer's site towards the Sun. Each ray is characterized by an incident direction 
$\bf{s_O}$ at the observer's site $\bf{x_O}$. The elementary surface for evaluating the irradiance
at $\bf{x_O}$ is oriented with its normal vector $\bf{n}_{\bf{x_O}}$ following the line that joins
$\bf{x_O}$ and the planet's centre. 
}
%/LIRIS2008/SKETCH0/
\end{figure*}

Figure (\ref{sketch0_fig}) sketches
how the refraction-bent rays are traced sunwards from the 
observer's site and, in turn, how 
Eq. (\ref{irradianceO_eq}) is evaluated for the direct sunlight component. 
In our numerical implementation of the integral, 
one ray is traced for each specified discrete element of 
$\bf{s_O}\cdot n_{\bf{x_O}}$$d\Omega(\bf{s_O})$. 
(We throughout refer to rays, although it would be more appropriate to refer 
to ray bundles.)
Since the integration is carried out at $\mathbf{x_O}$, we do not have to
correct for the optical phenomenon of attenuation by refraction 
\citep{link1969}. 
Attenuation by refraction is simply the 
reduction in the cross section (and therefore in the associated irradiance) 
of a ray bundle departing from the solar disk and reaching the observer's
site after crossing the Earth's atmosphere. % \citep{haysroble1968}. 
In other formulations that use the solar plane for integrating 
Eq. (\ref{irradianceO_eq}), this phenomenon must be explicitly 
introduced  as a factor weighing the size of the emitting 
solar disk parcel and its refracted image at the planet's terminator \citep{link1969}.  
This effect would be
especially important for rays crossing the lower layers of the atmosphere.
Those are also the rays more strongly affected by gas and aerosol extinction.

The formulation %sketched in {\S}\ref{theory_sec} 
holds valid for any combination of $e$ and $d_{\rm{{O}}}$, 
where $e$ is the geocentric angular distance from the area of lunar disk being probed to
the  geometrical umbra centre and $d_{\rm{{O}}}$ is the distance from
the observer's site on the Moon to the Earth's centre. 
 In the $d_{\rm{{O}}}$$\rightarrow$$\infty$ limit, the formulation
becomes relevant to the investigation of transiting extrasolar planets and
$e$ becomes a measure of the orbital phase. 
Numerically, we set $d_{\rm{{O}}}$$\rightarrow$$\infty$  by
assuming a sufficiently large $d_{\rm{{O}}}$. 
Special care is taken to ensure that all the formula occurring in the numerical
implementation adopt the correct forms in that limit case.

%\newpage
\section{\label{timevar_sec} The eclipse in progress}

We start by investigating the near-infrared spectra obtained at the WHT,
looking for changes in their structure during the umbra phase of the eclipse.
The exercise is useful to further validate our model of lunar eclipses  
and to show how the depth of molecular bands in the spectra varies as the
eclipse progresses.

\begin{figure*}[h]
%\epsscale{1.80}
\noindent\includegraphics[width=40pc]{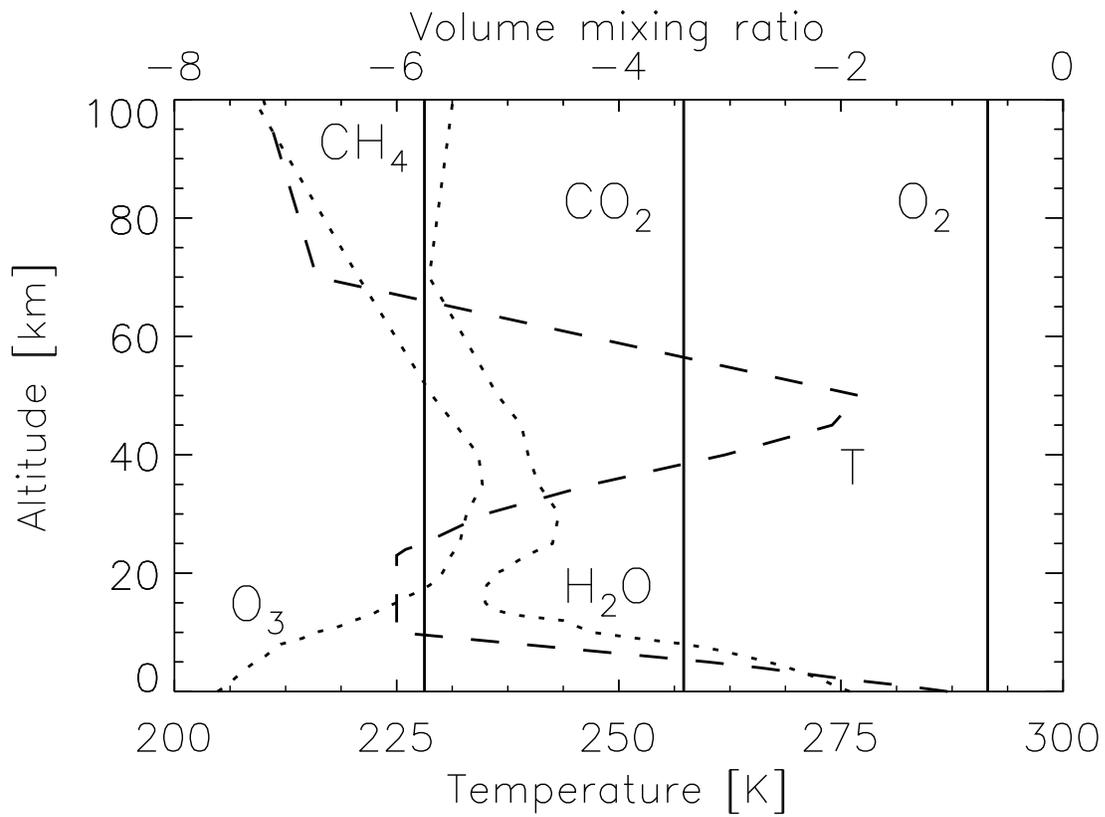}
\caption{\label{atmosp_fig} Vertical profiles of temperature (dashed curve) and
volume mixing ratios (solid and dotted curves) in the nominal atmosphere
}
%/LIRIS2008/ATMOSP/atmosp.pro
\end{figure*}

The nominal atmosphere in the simulations
takes the temperature, O$_3$ and H$_2$O (vapor) profiles from the
FSCATM subarctic summer model atmosphere \citep{galleryetal1983}, 
and constant volume mixing ratios of 0.2094, 3.833$\times$10$^{-4}$ and  
1.779$\times$10$^{-6}$ for O$_2$, CO$_2$ and CH$_4$ \citep{hansensato2004}, 
respectively. The vertical profiles of 
temperature and the volume mixing ratios are graphed in Fig. (\ref{atmosp_fig}).
The density at all altitudes was determined by integration of the hydrostatic balance equation.
For the Collision Induced Absorption (CIA) bands of oxygen at 1.06 and 1.27 $\mu$m, we proceeded
as follows.
For the O$_2$$\cdot$O$_2$ CIA band at 1.06 $\mu$m, 
we adopted the binary cross sections measured  
at 230 K in a 75/25\% mixture of O$_2$/N$_2$ by \cite{smithnewnham2000}.
For the O$_2$$\cdot$O$_2$ + O$_2$$\cdot$N$_2$ CIA band at 1.27 $\mu$m, 
we used the binary cross sections measured at 253 K by \citet{mateetal1999}. 
Both sets were rescaled to have peak values in air of 
1.7$\times$10$^{-45}$$\times$0.21=3.57$\times$10$^{-46}$ 
and 1.7$\times$10$^{-45}$$\times$0.21$\times$4=1.43$\times$10$^{-45}$ 
cm$^5$ molec$^{-2}$ at 1.06 and 1.27 $\mu$m, respectively.
The selected band shapes and scaling factors are consistent with the conclusions
drawn in a recent investigation 
of solar occultation data obtained with the SCIAMACHY spectrometer aboard ENVISAT
\citep{garciamunozbramstedt2012}. 
We evaluated the optical opacity 
for both CIA bands %at 1.06 and 1.27 $\mu$m
through $d\tau=\sigma_{\rm{CIA}}[\mbox{O}_2][\mbox{X}]ds$,
where $ds$ is the differential integration path, [$\cdot$] stands for number
density, X is the O$_2$ collision partner in the CIA band (O$_2$ at 1.06 $\mu$m
and O$_2$ + N$_2$ at 1.27 $\mu$m) and 
$\sigma_{\rm{CIA}}$ are the binary cross sections.
We set an opaque cloud layer with cloud tops at 6 km and an aerosol extinction 
profile at 1.02 $\mu$m about 4 times the September 2008 profile of \citet{siorisetal2010}. 
Further, we assumed that the aerosol optical properties are
wavelength-independent over the explored range of wavelengths. 
The nominal atmosphere is consistent with the findings by \citet{garciamunozetal2011}
from optical data of the same eclipse that point to a heavy aerosol
loading following the eruption of the Kasatochi volcano one week ahead of the
eclipse. 
Scattered sunlight is negligible longwards of 1 $\mu$m 
and therefore omitted. The simulated spectra were convolved with a 
Gaussian line shape at a resolving power of about 1,000.
 
A critical aspect in the comparison between observations and simulations is the 
solar elevation angle, $e$, or distance from the geometrical umbra centre to the 
projection of the instrument's collecting element on the lunar disk. 
No images were taken of the slit projected on the Moon and   
$e$ had to be estimated indirectly. For that purpose, we followed two different
methods.
Method 1) estimates $e$ from the telescope pointing coordinates recorded in the
FITS raw datafiles. 
The Sun, Earth and Moon coordinates, the coordinates at the ORM and 
the angular size of the lunar disk were taken from the JPL HORIZONS service 
\citep{giorginietal1996}. 
In method 2), we produced geocentric images of the lunar disk with the
GeoViz software\footnote{New Horizons Geoviz by 
H. Throop, http://soc.boulder.swri.edu/nhgv/}. From the Moon's centre, we 
displaced the slit south in equatorial declination on the image until only half 
of the 4.2-arcmin slit rested on the lunar disk. 
From the image, we recorded the coordinates of the mid-point projection of the
slit on the lunar disk.
Then, $e$ was calculated as the geocentric angle from that point to the 
umbra centre. 

\begin{figure*}[h]
%\epsscale{1.80}
\noindent\includegraphics[width=40pc]{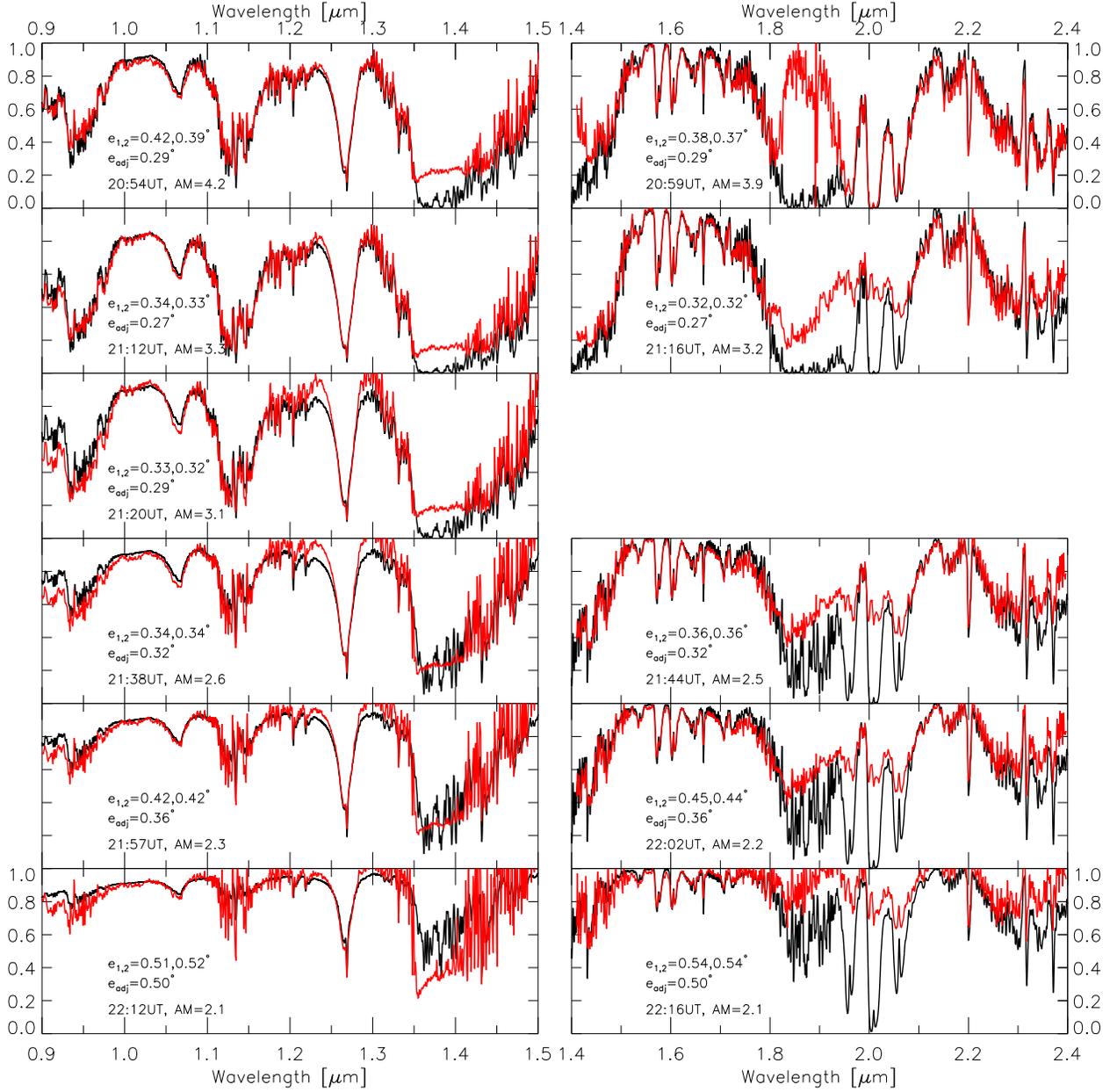}
\caption{\label{timevar_fig} 
In red, the eclipse spectra in $zj$ (left) and $hk$ (right). 
In black, the model simulations. The text inset shows the estimated 
$e$ angles and the $e$ angle adjusted in the model simulation to match optimally
the observed spectra. AM is the airmass for the Moon-to-telescope optical path.
}
%/LIRIS2008/TIMEVAR/timevar.pro
%/LIRIS2008/TIMEVAR/timevar_graph_v4.pro
\end{figure*}

Figure (\ref{timevar_fig}) shows the set of 11 umbra spectra together
with our model simulations. 
We arranged the $zj$ and $hk$ spectra in pairs according to the times of 
observation. Each pair contains data collected within an
interval of no more than 6 minutes. 
To first approximation, each pair can be seen as an uninterrupted spectrum 
from 0.9 to 2.4 $\mu$m. 
For the model simulations, we adjusted the $e$ parameter 
 to ensure an optimal match of the eclipse data. 
In some cases we modified the H$_2$O content by an amount that was 
always  less than 10\% of the nominal profile. 
The text in the graphs gives the time at mid-exposure and airmass for the 
Moon-to-telescope optical path, along with 
the estimated ($e_1$ and $e_2$) and adjusted ($e_{\rm{adj}}$) 
solar elevation angles. 
The quality of each match was judged by giving 
special weight to the O$_2$ and CIA bands in $zj$, and 
the CO$_2$ and CH$_4$ bands in $hk$. 
Absorption in both CIA bands remains linear throughout the eclipse. 
Indeed, we note the suitability of the CIA band at 1.06 $\mu$m, 
which occurs over a region only moderately affected 
by other molecular bands, for monitoring the Moon's progress across the umbra. 
Overall, the agreement between the eclipse data and the simulations is 
consistently good over the entire spectral range. 
Exceptions are the regions of strong water absorption at 1.35--1.45 and 
1.8--2.1 $\mu$m, where the telluric correction introduces obvious artifacts. 
A more flexible implementation of aerosol extinction in the simulations would
have surely improved the match between the observational and synthetic spectra,
but it was judged that such an effort was not critical for the present purpose.

The minimum $e_{\rm{adj}}$ occurs at 21:12--21:16 UT, 
in fair agreement with the prediction of geometrical greatest eclipse 
for 21:10:06UT.\footnote{Eclipse 
Predictions by F. Espenak, 
http://eclipse.gsfc.nasa.gov/eclipse.html}
The comparison between the estimated and adjusted angles 
indicates that $e_1$ and $e_2$ are systematically larger than $e_{\rm{adj}}$. 
The differences are typically
$\sim$0.05$^{\circ}$ except in the 20:54--20:59 UT spectrum, in which case it 
reaches $\sim$0.1$^{\circ}$. 
This spectrum was acquired at a rather high airmass, and refraction in the
Moon-to-telescope optical path may thus play a role. 
The 4.2--arcmin slit spans $\sim$1/15-th of the lunar disk diameter. 
The sunlight that enters the slit arrives from a range of distances to the umbra
centre that may lead to differences in $e$ of $\sim$0.04$^{\circ}$. 
Further, the non-uniform albedo of the Moon along the slit may move the
effective slit centre away from its geometrical location. 
It is thus difficult to define an equivalent solar elevation angle from purely 
geometrical considerations. 

In addition to the above arguments, one may also expect that local 
features in the atmosphere intercepted by the refracted sunlight rays will have an
impact on $e_{\rm{adj}}$. 
To explore this possibility, 
we conducted a few simulations in modified conditions of clouds and aerosols.
As a rule, setting the cloud tops below the 
nominal 6--km level leads to deeper absorption bands than in 
nominal conditions.
If, for instance, the cloud tops are set at 3 km, the 
21:12--21:16UT eclipse spectra are optimally reproduced with $e$$\sim$0.29$^{\circ}$, rather
than 0.27$^{\circ}$. 
The effect of low clouds on $e_{\rm{adj}}$ 
diminishes progressively as $e$ increases. 
The nominal aerosol profile extends well up to $\sim$16--17 km and results in
sunlight extinction over a broad range of altitudes. We verified that
reducing the aerosol extinction to the September 2008 levels 
of \citet{siorisetal2010} has a minor 
impact on the spectrum's structure near greatest eclipse.

\begin{figure*}[h]
\epsscale{1.00} 
\plotone{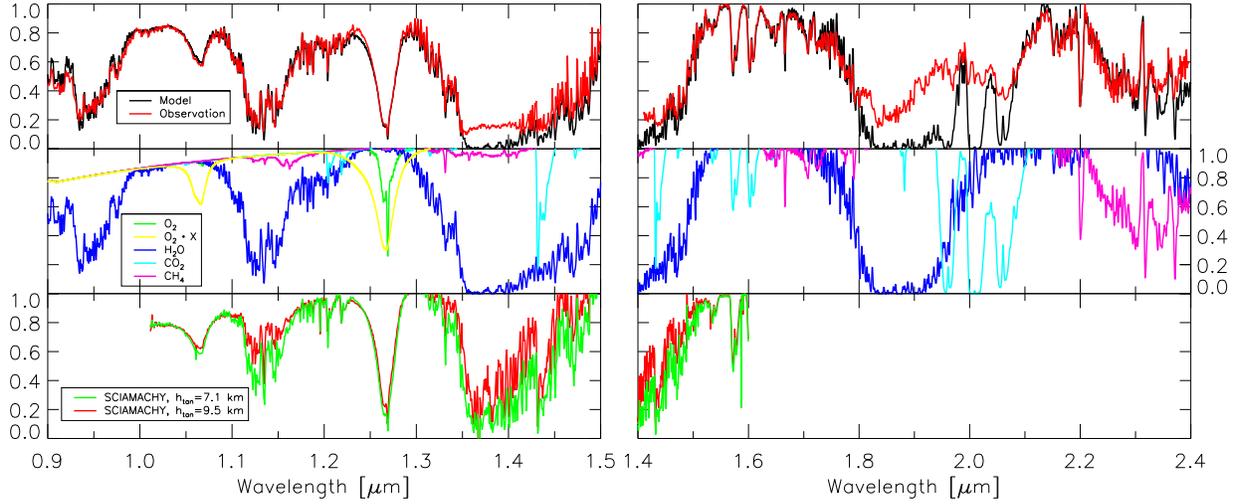} 
\caption{\label{timevar2_fig}
Top: Lunar eclipse spectrum at 21:12--21:16UT and model simulation. 
Middle: O$_2$, O$_2$$\cdot$X, H$_2$O, CO$_2$ and CH$_4$ contributions to the
model simulation. Bottom: SCIAMACHY solar occultation spectra measured 
from tangent altitudes of 7.1 and 9.5 km.
}
%/LIRIS2008/TIMEVAR/timevar2_graph_v4.pro
\end{figure*}

Figure (\ref{timevar2_fig}, top) shows the 21:12--21:16UT eclipse
spectrum and the corresponding simulation. 
In Fig. (\ref{timevar2_fig}, middle), the color lines are the contributions from
H$_2$O, O$_2$, CO$_2$, CH$_4$ and the CIA bands of oxygen to the simulation. 
For $e$$\sim$0.3$^{\circ}$, the sunlight rays that reach 
the area of lunar disk being probed 
have their closest approach to the Earth's surface at altitudes 
in the range $\sim$2--11 km \citep{garciamunozpalle2011}. 
The range is effectively narrower %in our simulations 
because the lower altitudes are optically thick and, in addition, we
assumed opaque clouds below 6 km. 
The spectrum of Fig. (\ref{timevar2_fig}, top) 
is indeed well approximated by a limb-viewing transmission spectrum 
of the atmosphere as seen from a tangent altitude of $\sim$10 km. 
For comparison, Fig. (\ref{timevar2_fig}, bottom) shows two solar occultation 
spectra measured with the SCIAMACHY spectrometer 
for tangent altitudes of $\sim$7.1 and 9.5 km. 
The resemblance of the lunar eclipse spectrum to the SCIAMACHY
spectra is apparent.

\begin{figure*}[h]
%\epsscale{1.80}
\noindent\includegraphics[width=20pc]{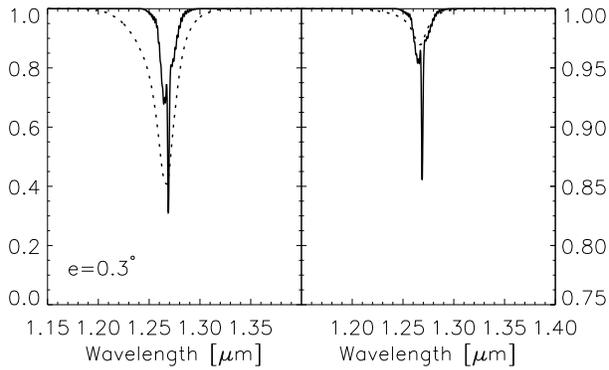}
\caption{\label{timevar3_fig} 
The O$_2$ (solid) and CIA (dotted) absorption bands at 1.27 $\mu$m for 
$e$ angles of 0.3 and 0.7$^{\circ}$. Far from the umbra centre, the 
CIA band loses relative importance with respect to the monomer band.
%for e=0.3, O2=54.7 A, Dim=181.6 A, ratio=3.31;
%for e=0.7deg, O2=8.62, Dim=7.79, ratio=0.904
}
%/LIRIS2008/TIMEVAR/timevar3_graph.pro
\end{figure*}

For $e$$\sim$0.5$^{\circ}$ (still in the umbra), the range of altitudes of closest
approach is $\sim$4--16 km, and 
for  $e$$\sim$0.7$^{\circ}$ (at the
umbra/penumbra edge) it is $\sim$8--65 km.
The shift and widening of the range of altitudes for closest approach 
has notable consequences on the spectrum's structure. 
For molecules whose densities decay monotonically with altitude, increasing
$e$ means the dilution in the spectrum of the molecular signature.
This is due, first, to a diminished contribution from the
atmosphere's   optically
thicker layers and, second, to a larger amount of light passing unattenuated
through the upper altitudes. 
This effect is especially apparent for H$_2$O and the 
O$_2$$\cdot$O$_2$ + O$_2$$\cdot$N$_2$ collision complex 
because their densities drop at a faster rate than those of
(nearly) well-mixed molecules such as O$_2$, CO$_2$ and CH$_4$. 
The simulations of the monomer and CIA bands of oxygen at 1.27 $\mu$m 
in Fig. (\ref{timevar3_fig}) illustrate this trend. 
For $e$=0.3$^{\circ}$, the O$_2$:CIA equivalent widths are in a
ratio of 1:3.3, whereas for $e$=0.7$^{\circ}$, the ratio is only 1:0.9.
Thus, as the observer probes regions of the lunar disk closer to the penumbra,
the relative contributions of H$_2$O and the collision complex 
with respect to the other molecules tend to decrease.

\subsection{The impact of the upper atmosphere on umbra spectra}

The umbra spectrum published 
by \citet{palleetal2009} with data of the 16 August 2008 event 
is representative of conditions near greatest eclipse
\citep{garciamunozetal2011} and the atmosphere at mid-to-north latitudes over
the Atlantic. At the relevant 
$e$$\sim$0.3$^{\circ}$, the absorption features are formed in refracted sunlight ray
trajectories 
that approach the Earth's surface at minimum distances $\lesssim$11 km. 
The occurrence of strong H$_2$O and CIA band features confirms
the importance of the lower altitudes. 

\citet{palleetal2009} mention the %tentative 
identification of the Na {\small{I}}  neutral atom %and the Ca {\small{II}} ion 
at optical wavelengths in their published spectrum. 
The eclipse was contemporary with the meteor shower of the Perseids, which means
that abnormally elevated amounts of that
metal in the mesosphere and lower thermosphere  might be expected
\citep{plane2003}. 
Following \citet{fussenetal2004}, we estimate that 
in globally-averaged conditions the optical thickness 
at the resolution of the umbra spectrum for the 
Na {\small{I}}  doublet at 0.589 $\mu$m from 
a tangent altitude of 11 km  is  $\sim$2$\times$10$^{-3}$.
The sodium peak density is known to  increase sporadically 
by factors of up to 10
in layers a few kilometer thick (which entails an increase in the integrated
column by a factor of a few), 
sometimes over horizontal spans of hundreds of kilometers, 
especially during meteor showers
 \citep{moussaouietal2010,plane2003,douetal2009}. 
Even then, such a weak signature makes the detection of the Na {\small{I}}
neutral atom
challenging for a spectrum of moderate spectral resolution. 
Thus, our analysis indicates that the
identification of the Na {\small{I}} doublet in the \citet{palleetal2009} spectrum is
questionable and requires additional confirmation.

%\newpage 
\section{The Earth in transit \label{transit_sec}} 
\subsection{The Earth-Sun system at mid-transit}

Everyday, sunsets offer real-life demonstrations
of the refraction of sunlight in the atmosphere. 
Looking from the ground over the horizon, 
the setting Sun appears about one solar diameter 
above its true elevation. 
Sunlight rays crossing the atmosphere and having their closest approach to the
Earth's surface  at a tangent altitude $h_{\rm{tan}}$ 
are approximately refracted by an angle 
$\alpha_{\rm{refr}}$$\sim$$(n(h_{\rm{tan}})$$-$$1)$$\times$$(2\pi R_p/H)^{1/2}$ 
\citep{baumcode1953,seagersasselov2000}. For a terrestrial radius $R_p$$\sim$6377 km, 
a constant scale height $H$$\sim$8 km 
and a typical refractivity at sea level in the optical
 $n(h_{\rm{tan}}$=$0)$$-$$1$$\sim$2.8$\times$10$^{-4}$, 
 the end-to-end refracted angle estimated from the above equation  
 is $\alpha_{\rm{refr}}$$\sim$1.14$^{\circ}$. 
Thus, over a distance of 1 AU, a ray incident on the planet 
that grazes the Earth's surface 
is deflected by $\sim$4.26 solar radii. 
Tracing the ray trajectories from a remote observer's site back towards 
the solar disk, one concludes that the mid-transit solar image at the 
observer's site is formed by 
sunlight rays having their closest approach to the planet's surface 
 at densities $\sim$1/4.26 or less than at sea level. 
Further, our estimate suggests that, at mid-transit, 
refraction prevents the remote observer's access to altitudes within a 
$z_{\rm{refr}}$-sized ring above the planet's surface. 
The exact size of this refraction-exclusion ring 
can only be determined by integrating the ray trajectories through the atmosphere
after prescribing the altitude-dependent refractivity of the atmospheric gas. 
Assuming, for the purpose of obtaining a first estimate, 
that the atmosphere is isothermal and that the refractivity decays with altitude
as $\exp{(-z/H)}$, one obtains that 
$z_{\rm{refr}}$$\sim$$H$$\ln{4.26}$$\sim$11.6 km. This 
approximate result is reached by equating the deflected distance of rays tangent
at an altitude $z_{\rm{refr}}$ and the solar radius, \textit{i.e.} 
$R_{\sun}= ( n(h_{\rm{tan}}=0)-1)\exp{(-z_{\rm{refr}}/H)} (2\pi R_p/H)^{1/2} a_{\sun}$. 
Here, $R_{\sun}$ and $a_{\sun}$ stand for solar radius and the planet's orbital
distance, respectively.

Our model simulations 
with the formulation of {\S}\ref{theory_sec} 
in the $d_{\rm{{O}}}$$\rightarrow$$\infty$ limit
provide refined magnitudes of the above estimates for $z_{\rm{refr}}$.
For the nominal atmosphere described above, a sunlight ray grazing the Earth's surface is 
refracted by $\sim$1.08$^{\circ}$ and only altitudes 
above $z_{\rm{refr}}$$\sim$13.2 km contribute to the mid-transit signal measured 
at the observer's site. 
Thus, the mid-transit spectrum measured by the remote observer 
will lack the signature of the atmospheric layers 
containing the bulk of the spectroscopically-active gases. 
The value of 
$z_{\rm{refr}}$ depends on the fluctuations in the 
atmosphere's density profile and water content through the refractivity of the
gas.
A sensitivity analysis perturbing those two parameters within the values
in the model atmospheres of \citet{galleryetal1983}
shows that $z_{\rm{refr}}$ typically lies between 12 and 14 km.

Figure (\ref{transit_fig}) displays in black the mid-transit spectrum of the Earth in two levels 
of approximation, namely with and without refraction, and three 
atmospheric scenarios. The scenarios represent: (1) a Rayleigh
atmosphere, free of clouds and aerosols;
(2) an atmosphere with cloud tops at 2 km, the
background aerosol extinction profile for September 2007 at about 1 $\mu$m 
published by \citet{siorisetal2010} multiplied by (1.02/$\lambda$
[$\mu$m])$^{1.2}$ at other wavelengths; 
and, 
(3) the aerosol-rich, cloudy atmosphere described in {\S}\ref{timevar_sec}. 
The aerosol loading in scenario (2) might be seen as a 
plausible representation of globally-averaged conditions  \citep{hayashidahorikawa2001}. 
The aerosol loading in scenario (3) is abnormally elevated for typical Earth
conditions but may be representative of the atmosphere after a major volcanic
eruption \citep{garciamunozetal2011}. 
We note, however, that in our own Solar System, 
Mars undergoes episodic events of 
abnormally elevated aerosol amounts being transported across the globe and thus
affecting the aerosol loading on global scales, %\citep{rafkin2012}, 
and that Venus is enshrouded by a complex system of clouds and haze layers up to
pressures well below 1 atm.

We use the equivalent height, $h_{\rm{eq}}$, defined according to:
\begin{equation}
\frac{F_p}{F_{\odot}}=1- \left( \frac{R_p+h_{\rm{eq}}}{R_{\odot}} \right)^2,
\label{heq_eq}
\end{equation} 
to provide a measure of the atmosphere's thickness
opaque to the incident sunlight. 
As a consistency check, the refractionless calculations %using the formulation of {\S}\ref{theory_sec} 
were tested against 
the more usual formulation for the in-transit stellar dimming \citep{hubbardetal2001}:
\begin{equation}
\frac{F_p}{F_{\odot}}= \frac{2\pi \int_{R_p}^{R_{\odot}} \exp{(-\tau(r_b))} r_b
dr_b}{\pi R^2_{\odot}}
\end{equation}
with:
\begin{equation}
\tau(r_b)=  \int_{r_b}^{R_{\odot}}  \frac{ 2 r_b \gamma(r) r/r_b
}{\sqrt{(r/r_b)^2-1}}d(r/r_b),
\label{tau_eq}
\end{equation}
where $\tau(r_b)$ is the optical thickness at an impact altitude $r_b-R_p$ and
$\gamma(r)$ is the optical extinction coefficient at altitude $r-R_p$ above the
planet's surface. 
The agreement between the two implementations proved to be excellent. 
It must be mentioned that Eq. (\ref{tau_eq}) was integrated 
in $t$, =$\cosh^{-1}{(r/r_b)}$, 
rather than in $r/r_b$, to
avoid the singularity in the denominator of the integrand. 

Figure (\ref{transit_fig}) clearly shows how refraction removes much of the
mid-transit spectrum's structure. An exception to that 
is the Chappuis band of ozone, which absorbs noticeably in all cases 
shortwards of $\sim$0.8 $\mu$m. 
Terrestrial ozone densities peak in the stratosphere, and 
ozone in the refraction-exclusion ring contributes in a minor way 
to the disk-integrated signature. 
The features longwards of 1 $\mu$m 
are readily
identifiable by comparison with Fig. (\ref{timevar2_fig}). 
Shortwards of 1 $\mu$m, 
the sharp features at 0.76, 0.69 and 0.63 $\mu$m are the
$X$($v''$= 0)$\rightarrow$$b$($v'$= 0, 1, 2) bands of O$_2$, respectively. 
The remaining structure in that region is largely due to H$_2$O. 
With respect to the nearby continuum, the equivalent heights of the
discrete features in the spectra of the refractive atmosphere 
are always less than $\sim$10 km in the 0.4--2.5 $\mu$m spectral range. 
In going from atmospheric scenarios (1) to (3) it is seen that 
refraction becomes less important 
as aerosol extinction near the $z_{\rm{refr}}$ boundary becomes dominant. 
The visual inspection of the transmission spectra published by 
\citet{ehrenreichetal2006} and \citet{kalteneggertraub2009} for an Earth twin
shows that their spectra are generally consistent with
our refractionless calculations.

Were the Sun and Earth observed at mid-transit from a remote distance, the 
transmitted sunlight reaching the remote observer's site would be determined 
by sunlight photons crossing the Earth's atmosphere at 
altitudes above $\sim$12--14 km. 
As seen in Fig. (\ref{transit_fig}), ignoring the refractive nature of the Earth's
atmosphere would mean a significant overestimation of most spectral features. 
The overestimation is more obvious when the model atmosphere is assumed 
free of clouds and aerosols since airborne particles 
act as natural barriers to the transmission of
light through the lower altitudes.
Because meteorological activity is largely confined to 
the troposphere, the planet's meteorology will have limited impact on the  
in-transit transmission spectrum. 

The sunlight rays passing closer to the 
planet's surface at mid-transit originate from the near-limb region of the opposite solar hemisphere. 
In order to assess the impact of solar limb darkening, we repeated the calculation for the 
three scenarios described above including a linear limb-darkening law for the solar 
brightness, 
$U=1-u_1(1-\mu_{\odot})$, and $u_1$=0.6 \citep{gimenez2006}. To proceed with the
comparison, we redefined $h_{\rm{eq}}$ in the way:
$$
\frac{F_p}{F_{\odot}}=1- \frac{3}{3-u_1}  \left( \frac{R_p+h_{\rm{eq}}}{R_{\odot}} \right)^2,
$$
to account for the $(3-u_1)/3$ factor in the limb-darkened solar irradiance,
$F_{\odot}$. Our model simulations indicate that $h_{\rm{eq}}$ differs by less than
$\sim$1 km in the cases with $u_1$= 0 and 0.6. 
Thus, solar limb darkening affects in a negligible way 
the spectrum's structure at mid-transit.

As in {\S}\ref{timevar_sec}, we calculated the equivalent widths of the monomer
 and CIA bands of oxygen at 1.27 $\mu$m. 
For the refractive calculations of scenarios (1)--(3), we obtained that the 
O$_2$:CIA equivalent widths are in ratios of 
1:0.70, 1:0.64 and 1:0.47, respectively. 
The ratios indicate that the distinct identification of the CIA band
becomes more and more difficult as the atmospheric opacity of aerosols and
clouds increases.

\citet{garciamunozpalle2011} estimated that the sunlight scattered at the Earth's
terminator towards the eclipsed Moon could amount to up to $\sim$10$^{-7}$--10$^{-6}$ 
of the solar irradiance depending on the forward-scattering efficiency of the
airborne aerosols. 
During the eclipse, the magnitude of scattered sunlight depends on 
the ratio of the solid angles subtended by the Earth, 
$\sim$$1/d^2_{\rm{O}}$, and the Sun, $\sim$$1/(d_{\rm{O}}+a_{\odot})^2$, from
the lunar observer's site on the Moon. In the lunar eclipse, 
we used for $d_{\rm{O}}$ and $a_{\odot}$, 382665.9 km and 1 Astronomical Unit, 
respectively. 
Thus, the amount of stellar light scattered towards
the remote observer during the transit is 
$\sim$(10$^{-7}$--10$^{-6}$)$\times$$(d_{\rm{O}}+a_{\odot})^2$/$d^2_{\rm{O}}$$\sim$
10$^{-12}$--10$^{-11}$, which is much less than the $\sim$10$^{-4}$ dimming caused by the planet's
core or the 10$^{-7}$--10$^{-6}$ dimming attributable to 
the strongest atmospheric features. 
 
Next, we address under what conditions a lunar eclipse 
spectrum may be representative of the in-transit transmission spectrum of the 
remotely-observed Earth-Sun system. 
Since at mid-transit only altitudes above $\sim$12--14 km are probed, 
the eclipse phases matching more closely that condition entail solar
elevation angles $e$$\gtrsim$0.9$^{\circ}$, which are well within the penumbra
\citep{garciamunozpalle2011}. 
Thus, spectra from the penumbra, rather than from the umbra, will be better
representatives of the in-transit transmission spectrum for the
Earth-Sun system.
Figure 3 in \citet{garciamunozpalle2011} shows an eclipse 
spectrum at different phases. The comparison of that figure and our  
Fig. (\ref{transit_fig}) confirms that the eclipse spectrum tends to resemble
the mid-transit transmission spectrum as the eclipse approaches the umbra/penumbra
edge.

%\newpage
\subsection{The Earth-Sun system near ingress/egress}

We have so far focused on the spectrum at mid-transit, noting 
the effect of refraction on the equivalent height of the Earth's atmosphere
with respect to the situation for a virtual non-refractive atmosphere.
The refraction-bending of sunlight rays does, however, introduce phase-dependent
effects on the intensity and spectral structure of the transmitted sunlight 
that are worth exploring.

Out of transit,
refraction causes the brightening of the planet's hemisphere more distant
from the solar disk. This effect has been reported 
for Venus transits and causes a transient
halo at the planet's limb that is readily discernible from Earth
with the aid of a small telescope
between internal and external contacts. 
It is attributed to Lomonosov back in the 18th century 
the correct interpretation of this halo, a fact that, in turn, meant the first
evidence for an atmosphere at the planet \citep{link1959}. 
The last Venus transit occurred in June 2004 and provided numerous images of 
the halo that confirmed the pole-to-equator asymmetry in the optical properties
of the planet's atmosphere \citep{pasachoffetal2011,tangaetal2012}. 
The contribution of the halo to the net brightness of the planet-star system is
tiny for Venus 
but might amount to a detectable magnitude for transiting 
Jovian extrasolar planets on long-duration orbits \citep{sidissari2010}. 

The halo is a visual representation of the enhancement in the amount of 
refracted sunlight at the planet's outer hemisphere. 
This enhancement occurs together with a diminishment on the opposite
hemisphere. Figure (\ref{sketch_fig}) sketches the situation.
Sunlight rays passing through the outer hemisphere may be refracted by
an angle of up to $\sim$2$R_{\sun}/a_{\sun}$, whereas on the inner hemisphere
refraction angles of more than $\sim$2$R_p/a_{\sun}$ do not contribute
to the signal collected by the distant observer. 
Different refraction angles also mean different depths of the sunlight rays
into the atmosphere
and, in turn, a refraction ring whose size is larger on the inner hemisphere
than on the outer one.

The upper curves in Fig. (\ref{transit_refr_fig}) represent the equivalent
heights of the atmosphere from mid-transit ($e$=0$^{\circ}$) to 
internal contact ($e$$\approx$0.2641$^{\circ}$). We
omitted limb darkening to emphasize the purely refractive effects, an 
assumption that 
enables us to use Eq. (\ref{heq_eq}) as a valid definition for the effective
size of the atmosphere. The calculations assumed the Rayleigh atmosphere described
 above.
The main conclusion to draw from that set of curves 
is that the Earth's optical size varies with the phase of the transit 
as a result of the refraction of sunlight at the planet's limb. The Earth
appears larger near internal contact than at mid-transit because in the
proximity of the solar limb the optical enlargement of the inner hemisphere 
dominates over the shrinking of the outer one. The overall change in 
$F_p/F_{\sun}$ from mid-transit to internal contact amounts to about 3$\times$10$^{-7}$
for a variation in the equivalent height of 12 km. The detection of such a
dimming in the stellar light appears extremely challenging, 
especially after considering that most of the drop occurs near
ingress/egress where limb darkening effects on the planet-star 
lightcurve are more prominent. 
The upper set of curves in Fig. (\ref{transit_refr_fig}) also reveals that
all molecular band features tend to weaken for phases that move away from
mid-transit.

The lower set of curves in Fig$.$ (\ref{transit_refr_fig}) represent a measure
of the halo brightness from external contact ($e$$\approx$0.2690$^{\circ}$) 
to phases well out of the eclipse. 
We use an equivalent halo thickness defined through:
\begin{equation}
\frac{F_p}{F_{\odot}}=1 + 
\frac{2\pi R_p h_{\rm{halo}}}{\pi R^2_{\odot}}, 
\label{hhalo_eq}
\end{equation} 
that gives the thickness of a ring around the planet of the same brightness as
the Sun for $\mu_{\sun}$=1. 
The effective halo thickness is about 10 km at external 
contact and decays as the planet moves farther out of transit. 
The net brightening at external contact is about 3$\times$10$^{-7}$ and, again,
probably too faint for detection with the currently existing technology.

%\newpage
\subsection{An Earth in transit in a different pla\-netary system}

\begin{figure*}[h]
\epsscale{1.00}
%\noindent\includegraphics[width=20pc]{./FIGURES/transit_woLD.eps}
\plotone{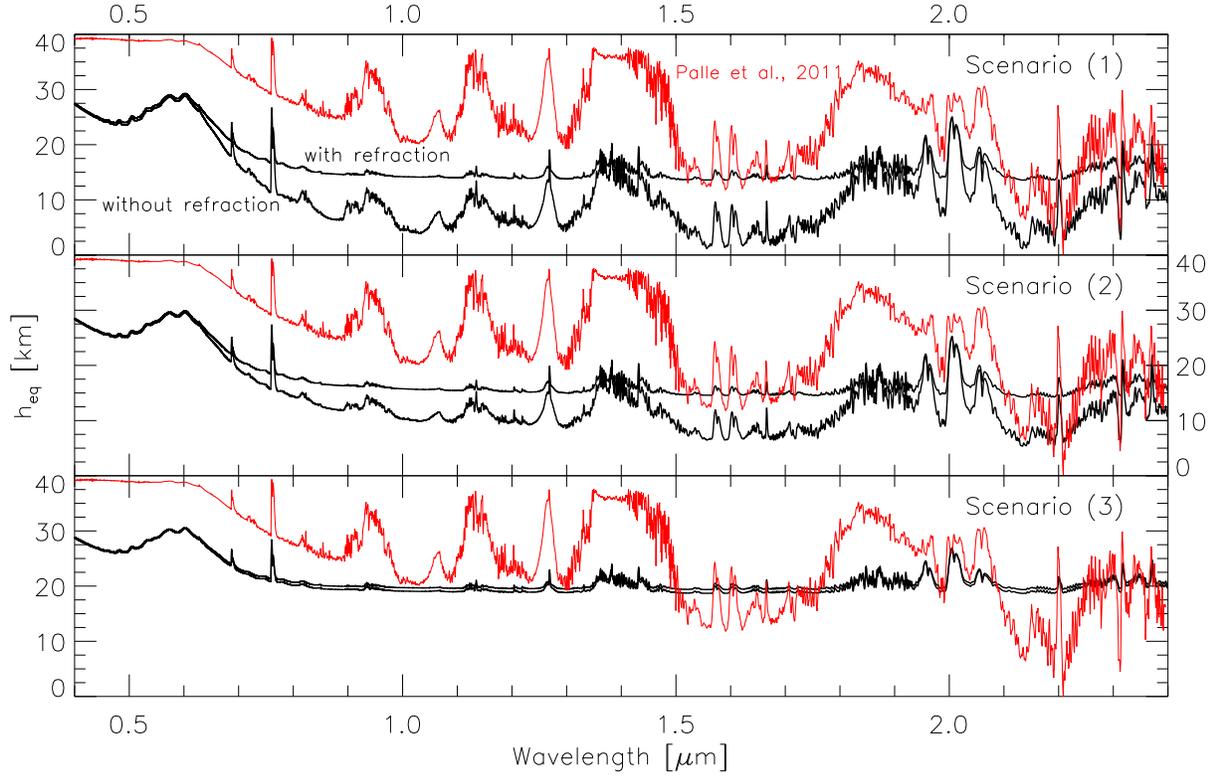}
\caption{\label{transit_fig} 
Equivalent heights for a refractive and a non-refractive 
atmosphere. Top panel: Rayleigh atmosphere. Middle panel: Globally-averaged conditions of
clouds and aerosols. Bottom panel: Aerosol-rich, cloudy atmosphere as described in 
{\S}\ref{timevar_sec}.
The red curve on each panel is 
the equivalent height as defined through the noise-free version of
Eq. 4 in \citet{palleetal2011}: 
$h_{\rm{eq}}=h_{\rm{TOA}}\times(1-\mathbb{T}_p)$ with 
$h_{\rm{TOA}}$=40 km and where $\mathbb{T}_p$ is the umbra spectrum published
by \citet{palleetal2009} normalized at its maximum near 2.2 $\mu$m.
}
%/LIRIS2008/TRANSIT/transit_graph_v4.pro
\end{figure*}

\begin{figure*}[h]
\epsscale{1.00}
%\noindent\includegraphics[width=20pc]{./FIGURES/sketch_fig.eps}
\plotone{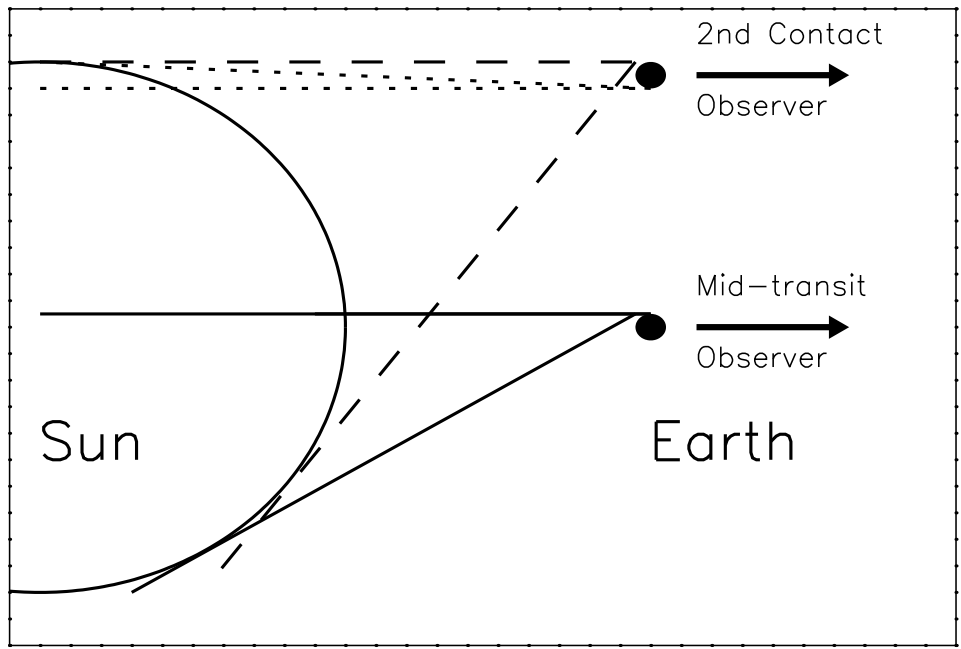}
\caption{\label{sketch_fig} 
The Earth-Sun system at mid-transit and at 2nd contact. In each case, the 
sunlight that traverses the Earth's limb originates from a different zone 
of the solar disk. At 2nd contact, the outer hemisphere sees the entire solar
disk, whereas the inner hemisphere sees only a narrow zone close to the Sun's
edge.
}
%/LIRIS2008/SKETCH/sketch.pro
\end{figure*}

\begin{figure*}[h]
\epsscale{1.00}
%\noindent\includegraphics[width=20pc]{./FIGURES/transit.eps}
\plotone{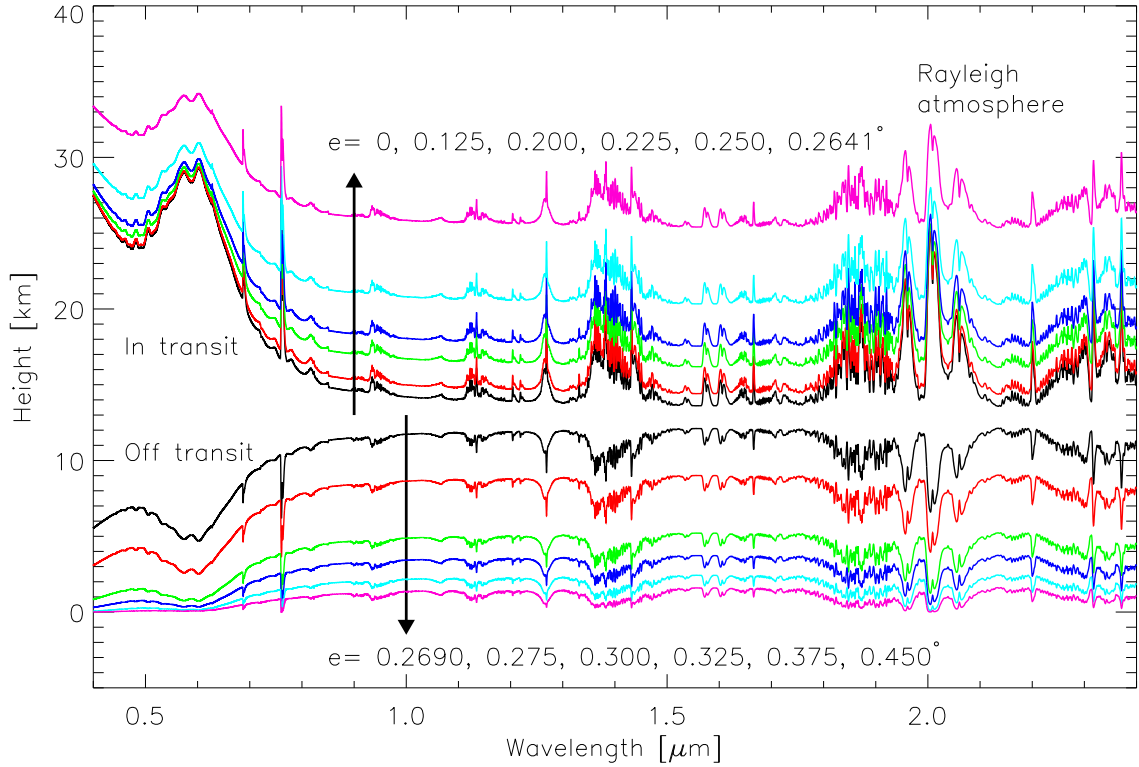}
\caption{\label{transit_refr_fig} 
Equivalent heights for the Earth-Sun system in transit (upper set of curves) and
halo heights for the system off transit (lower set of curves) at various phases ($e$). 
Limb darkening has
been omitted to highlight the impact of refraction. For the in-transit curves,
h$_{\rm{eq}}$ is defined through Eq. (\ref{heq_eq}), whereas for the off-transit
curves, h$_{\rm{eq}}$ is defined through Eq. (\ref{hhalo_eq}). The
calculations in this figure are specific to a Rayleigh atmosphere.
}
%/LIRIS2008/TRANSIT/transit_graph_vse.pro
\end{figure*}

We have seen that at mid-transit refraction imposes a ring around the Earth's terminator 
that deflects some line of sights from courses reaching the solar disk or,
equivalently, some sunlight rays incident on the planet 
from courses reaching the remote observer. 
The size of this refraction-exclusion ring depends on the 
refractive properties of the atmosphere but also 
on the stellar radius and the planet's orbital distance. 
In standard conditions of temperature and pressure, 
the N$_2$ and O$_2$ refractivities are about 1.6 times less
than the CO$_2$ refractivity but about twice the H$_2$ refractivity.
For a virtual Earth twin whose atmospheric properties remained unchanged when 
placed in a planetary system other than ours, 
the ring will shrink if the planet is 
 closer than 1 AU to the star  ($a_s < a_{\odot}$) 
or if the host star is larger than one solar radius ($R_s > R_{\odot}$). 
The inequality $\alpha_{\rm{refr}} a_s / R_s$$>$ 1 for the ratio of 
the refraction-deflected distance,
$\alpha_{\rm{refr}} a_s$ (with $\alpha_{\rm{refr}}$$\sim$1$^{\circ}$), 
and the stellar radius establishes the
approximate geometrical condition for which a refraction-exclusion ring exists
at sub-atmospheric pressures. 
Speculations on possible values for $\alpha_{\rm{refr}} a_s / R_s$ 
find their justification in the wide variety of known planetary systems and the
even wider variety that is likely to be discovered.
As a matter of fact, Fig. (\ref{transit_fig}) applies to a broader range of $a_s/a_{\odot}$ 
and $R_s/R_{\odot}$ parameters than first stated. 
The lower synthetic curve represents approximately the condition 
$a_s/a_{\odot}$$\times$$R_{\odot}/R_s$$<$1 and small enough so that refraction
has a minor impact on the in-transit transmission spectrum.
In turn, the upper synthetic curve
represents the condition $a_s/a_{\odot}$$\times$$R_{\odot}/R_s$$\sim$1. 
Those two curves bracket all possible conditions for 
the intermediate values of $a_s/a_{\odot}$$\times$$R_{\odot}/R_s$. 

Oxygen is a potential biomarker and therefore constitutes
a particularly appealling target for future searches.
From our Fig. (\ref{transit_fig}), it is
seen that the O$_2$ $X$(0)$\rightarrow$$b$(0) band at 0.76 $\mu$m may
be the best candidate for an oxygen search, although the CIA bands at
1.06 and 1.27 $\mu$m (the latter blended with the simultaneous 
monomer band) are also good candidates when the
lowermost layers of the in-transit extrasolar planet are probed. 

Although refraction will generally pose an additional challenge to the
characterization of Earth-sized planets, the added 
difficulty may be minor in some configurations of interest. 
That is indeed the case for an Earth twin orbiting within the habitable
zone (HZ) of some M-type stars, 
a planet-star combination that yields relatively favourable areal ratios.
The HZ is the distance around a star in which 
a planet's water might be expected to exist in liquid state. 
Taking for the HZ the reference orbital distance quoted by 
\citet{kalteneggertraub2009}, 
$a_{s}/a_{\odot}= (T_{\rm{s}}/T_{\odot})^2  (R_{\rm{s}}/R_{\odot})$, where,
following our earlier convention,  
subscripts $_{\rm{s}}$ and $_{\odot}$ denote stellar and solar magnitudes, 
respectively, one arrives at the condition, independent of $R_s$, 
$(T_s/T_{\odot})^2 < (R_{\odot}/a_{\odot})$$/\alpha_{\rm{refr}}$ for the stellar
effective temperature below which refraction effects are minor. 
If we adopt $\alpha_{\rm{refr}}$= 1.08$^{\circ}$ at the 1-atm pressure level, 
the condition becomes $T_s/T_{\odot}$$\le$0.5. 
Thus, an Earth twin that 
orbited within the HZ of an M-type star of effective temperature about 
2900 K would produce mid-transit 
spectra with a closer resemblance to the refractionless simulations 
in Fig. (\ref{transit_fig}) than to the refractive simulation of the Earth-Sun system. 
Obviously, other intermediate situations would be possible
for $T_{\rm{s}}$ values in between 2900 and 5800 K. 

Discussing detectability issues for specific telescope sizes in the way it is
done by \citet{ehrenreichetal2006} or  \citet{kalteneggertraub2009}
is beyond the scope of our work. 
\citet{kalteneggertraub2009} do recognize the potential impact of refraction 
on the lightcurves of Earth-sized planets. 
They do not, however, give the size of the refraction
ring in their investigation of the in-transit transmission spectrum of an Earth
twin orbiting M-type stars nor explain the way refraction is implemented. 
The authors mention an altitude of 6 km below which the terrestrial
atmosphere is opaque to grazing sunlight rays
but do not state how much of that is contributed by refraction.
As pointed out above, the size of the refraction-exclusion ring 
depends on the refractive properties of the 
atmosphere but also on both the stellar radius and the planet's orbital distance. 
Since \citet{ehrenreichetal2006} 
account in some cases for the presence of clouds
at altitudes of up to 10 km, the conclusions drawn from their cloudy models of an
Earth twin might be appropriate for specific conditions in which refraction 
plays a significant role. 
 
\citet{palleetal2011} discussed the in-transit detectability of the strongest
absorption features from 0.4 to 2.4 $\mu$m using scaled versions of
the \citet{palleetal2009} umbra spectrum. 
Setting the noise terms to zero in Eq. 4 of \citet{palleetal2011}, 
the equivalent height proposed in that work can be written as 
$h_{\rm{eq}}=h_{\rm{TOA}}\times(1-\mathbb{T}_p)$, where $h_{\rm{TOA}}$ is a specified
top-of-the-atmosphere height and $\mathbb{T}_p$ is the umbra spectrum normalized
to one near 2.2 $\mu$m. Our Fig. (\ref{transit_fig}) shows in red 
the resulting equivalent heights for the specific case 
$h_{\rm{TOA}}=$40 km investigated in \citet{palleetal2011}. 
The comparison with the upper (black) synthetic curve in the top panel demonstrates 
that for an Earth-like extrasolar planet 
lacking aerosols and clouds and close enough to a
solar-type star so that refraction effects are minor,  
the approach followed by \citet{palleetal2011} overestimates moderately
%, but is reasonably consistent with, 
the model predictions for a Rayleigh atmosphere. 
This is because the umbra spectrum published by \citet{palleetal2009} 
results from probing preferentially the lowermost altitudes of the atmosphere
during the eclipse. 
In all other cases displayed in Fig. (\ref{transit_fig}), 
and more especially when refraction effects play a role,
the equivalent heights for $h_{\rm{TOA}}=$40 km in \citet{palleetal2011} 
are larger by up to an order of magnitude than the current model predictions. 
It is apparent that fainter features will impose tougher requirements on their
detectability.

%\newpage
\section{\label{summary_sec}Concluding remarks} 
   
We investigated lunar eclipses of the Earth and the transits of Earth-like
extrasolar planets in a comparative manner.  
The Moon in umbra receives unscattered sunlight that has been transmitted through 
a section of the terrestrial atmosphere dictated by refraction. 
The diffuse sunlight that arrives at the eclipsed Moon is scattered from the
entire terminator. 
During the transit of the Earth-Sun system, at mid-event,
the refracted sunlight is transmitted through an annular 
ring that fully encloses the planet. 
The refraction-bending of sunlight rays prevents sunlight rays 
passing through a refraction-exclusion ring above the planet's surface
from reaching the remote observer. For the Earth-Sun system, at mid-transit, 
this ring extends up to 12--14 km, thus blocking the observer's access to 
the denser atmospheric layers.
It may be stated that umbra eclipses maximize the contribution of the lowermost
atmosphere and, to some extent, of scattered sunlight, 
whereas during the transit of an Earth-like extrasolar planet 
the transmitted signal is dominated by
altitudes above a refraction-exclusion ring 
and scattered sunlight becomes negligible.

In the course of this work it has become apparent that a genuine transmission
spectrum of the Earth in transit as observed from a remote distance is more
difficult to obtain than what had been anticipated. Lunar eclipse spectra
obtained during the penumbra are more representative of the in-transit spectrum
than those obtained during the umbra. In the context of Earth-like extrasolar
planets, the lunar eclipse spectra are helpful to fine tune synthetic spectra
incorporating the fairly good knowledge of the atmosphere obtained over decades
of Earth's atmospheric research.

Refraction will affect the lightcurves of transiting Earth-like 
extrasolar planets. Its significance will depend
on the refractive properties of the planet's atmosphere, the stellar
size and the planet's orbital distance. 
All these factors must be considered 
in the future to predict and/or interpret the lightcurves of such planets.
The role of refraction will not be critical for Earth twins orbiting
within the HZ of M-type stars with effective temperatures of about 2900 K and
less. 
Finally, it is worth noting that 
recent lunar eclipse observations have contributed critically 
to bring up the importance of refraction in the future 
characterization of Earth-like extrasolar planets. 

\acknowledgments
The near-infrared spectra discussed in {\S}\ref{timevar_sec} were obtained 
with the William Herschel Telescope at the Spanish Observatorio del Roque de los
Muchachos of the Instituto de Astrof\'isica de Canarias. MRZO and PMR 
acknowledge support from projects AYA2010-21308-C03-02 and CGL2009-10641,
respectively.
Both MRZO and ELM acknowledge support from project AYA2011-30147-C03-03. 
ELM is grateful for financial support from RoPACS, 
a Marie Curie Initial Training Network funded by the
European Commission's Seventh Framework Program,
and by the CONSOLIDER-INGENIO GTC project. 
AGM acknowledges G. T. Fraser for providing 
the \citet{mateetal1999} binary 
cross sections and K. Bramstedt and 
the European Space Agency for the SCIAMACHY solar occultation spectra.

\clearpage

%% the file, uncomment the \plotone, \plottwo, and
%% If you need a layout that cannot be achieved with \plotone or
%% \plottwo, you can invoke the graphicx package directly with the
%% \includegraphics command or use \plotfiddle. For more information,
%% please see the tutorial on "Using Electronic Art with AASTeX" in the
%% documentation section at the AASTeX Web site,
%% http://www.journals.uchicago.edu/AAS/AASTeX.
%%
%% The examples below also include sample markup for submission of
%% supplemental electronic materials. As always, be sure to check
%% the instructions to authors for the journal you are submitting to
%% for specific submissions guidelines as they vary from

%% This example uses \plotone to include an EPS file scaled to
%% 80% of its natural size with \epsscale. Its caption
%% has been written to indicate that additional figure parts will be
%% available in the electronic journal.

%% Here we use \plottwo to present two versions of the same figure,
%% one in black and white for print the other in RGB color
%% for online presentation. Note that the caption indicates
%% that a color version of the figure will be available online.
%%

%% This figure uses \includegraphics to scale and rotate the still frame

%% If you are not including electonic art with your submission, you may
%% mark up your captions using the \figcaption command. See the
%% No more than seven \figcaption commands are allowed per page,
%% so if you have more than seven captions, insert a \clearpage

%% Tables should be submitted one per page, so put a \clearpage before

%% Two options are available to the author for producing tables:  the
%% deluxetable environment provided by the AASTeX package or the LaTeX
%% table environment.  Use of deluxetable is preferred.
%%

%% Three table samples follow, two marked up in the deluxetable environment,
%% one marked up as a LaTeX table.

%% In this first example, note that the \tabletypesize{}
%% command has been used to reduce the font size of the table.
%% We also use the \rotate command to rotate the table to
%% landscape orientation since it is very wide even at the
%% Note also that the \label command needs to be placed

%% This table also includes a table comment indicating that the full
%% version will be available in machine-readable format in the electronic

\clearpage

%% If you use the table environment, please indicate horizontal rules using
%% Do not put multiple tabular environments within a single table.
%% The optional \label should appear inside the \caption command.

%% If the table is more than one page long, the width of the table can vary
%% from page to page when the default \tablewidth is used, as below.  The
%% individual table widths for each page will be written to the log file; a
%% maximum tablewidth for the table can be computed from these values.
%% The \tablewidth argument can then be reset and the file reprocessed, so
%% that the table is of uniform width throughout. Try getting the widths
%% from the log file and changing the \tablewidth parameter to see how
%% adjusting this value affects table formatting.

%% The \dataset{} macro has also been applied to a few of the objects to
%% show how many observations can be tagged in a table.

%% Tables may also be prepared as separate files. See the accompanying
%% sample file table.tex for an example of an external table file.
%% To include an external file in your main document, use the \input
%% command. Uncomment the line below to include table.tex in this
%% sample file. (Note that you will need to comment out the \documentclass,
%% \begin{document}, and \end{document} commands from table.tex if you want
%% to include it in this document.)


%% The following command ends your manuscript. LaTeX will ignore any text


\begin{thebibliography}{100}

\bibitem[Batalha et al.(2011)]{batalhaetal2011}
Batalha, N. M., Borucki, W. J., Bryson, S. T., Buchhave, L. A., Caldwell, D. A.,
et al., 2011, \apj, 729, 27

\bibitem[Baum \& Code(1953)]{baumcode1953}
Baum, W.A., \& Code, A. D., 1953,  \aj, 58, 108

\bibitem[Berta et al.(2012)]{bertaetal2012} 
Berta, Z. K., 
Charbonneau, D., 
D\'esert, J.-M., 
Miller-Ricci Kempton, E.,
McCullough, P. R., et al., 2012,
\apj, 747, 35

Batalha, N. M., Borucki, W. J., Bryson, S. T., Buchhave, L. A., Caldwell, D. A.,
et al., 2011, \apj, 729, 27


\bibitem[Brown(2001)]{brown2001}
Brown, T. M., 2001, 
\apj, 553, 1006


\bibitem[Charbonneau et al.(2002)]{charbonneauetal2002}
Charbonneau, D., Brown, R. M., Noyes, R. W. \& Gilliland, R. L., 2002,
\apj, 568, 377

\bibitem[Charbonneau et al.(2009)]{charbonneauetal2009}
Charbonneau, D., Zachory, B. K., Irwin, J., Burke, C. J., Nutzman, P. et
al. (2009),
nature, 462, 891.


\bibitem[D\'esert et al.(2009)]{desertetal2009}
D\'esert, J., Lecavelier des Etangs, A., H\'ebrard, G., Sing, D. K., Ehrenreich,
et al., 2009,
Astron. \& Astrophys., 699, 478

\bibitem[Dou et al.(2009)]{douetal2009}
Dou, X.-K., Xue, X.-H.,  Chen, T.-D., Wan, W.-X., Cheng, X.-W.,
et al., 2009,
Ann. Geophys., 27, 2247

\bibitem[Ehrenreich et al.(2006)]{ehrenreichetal2006}
Ehrenreich, D., Tinetti, G., Lecavelier des Etangs, A., Vidal-Madjar, A. \&
Sensis, F., 2006,
Astron. \& Astrophys., 448, 379

%\bibitem[Ehrenreich et al.(2012)]{ehrenreichetal2012}
%Ehrenreich, D., Vidal-Madjar, A., Widemann, T., Gronoff, G., Tanga, P., 
%et al., 2012, 
%Astron. \& Astrophys., 537, L2

\bibitem[Fraser \& Lafferty(2001)]{fraserlafferty2001}
Fraser, G. T. \& Lafferty, W. J., 2001,
\jgr, 106, 31749

\bibitem[Fussen et al.(2004)]{fussenetal2004}
Fussen, D., Vanhellemont, F., Bingen, C., 
Kyr\"ol\"a, E., Tamminen, J., et al., 2004,
\grl, 31, L24110, doi:10.1029/2004GL021618

\bibitem[Gallery et al.(1983)]{galleryetal1983}
Gallery, W. O., Kneizys, F. X., and Clough, S. A., 1983, 
In: Environmental research paper ERP-828/AFGL-TR-83-0065, Hanscom.

\bibitem[Garc\'ia Mu\~noz \& Pall\'e(2011)]{garciamunozpalle2011} 
Garc\'ia Mu\~noz, A. \& Pall\'e, E., 2011, \jqsrt, 112, 1609

\bibitem[Garc\'ia Mu\~noz et al.(2011)]{garciamunozetal2011} 
Garc\'ia Mu\~noz, A. \& Pall\'e, E., Zapatero Osorio, M. R. \& Mart\'in, E. L., 2011,
\grl, 38, L14805

\bibitem[Garc\'ia Mu\~noz \& Bramstedt(2012)]{garciamunozbramstedt2012} 
Garc\'ia Mu\~noz, A. \& Bramstedt, K., 2012, \jqsrt, 113, 1566

\bibitem[Gim\'enez(2006)]{gimenez2006}
Gim\'enez, A., 2006, 
Astron. \& Astrophys., 450, 1231

\bibitem[Giorgini et al.(1996)]{giorginietal1996}
Giorgini, J. D., 
Yeomans, D. K., 
Chamberlin, A. B., 
Chodas, P. W., 
Jacobson, R.A., 
et al., 1996,
\baas, 28, 1158. 
%, "JPL's On-Line Solar System Data Service", 
%Bulletin of the American
Astronomical Society, Vol 28, No. 3, p. 1158, 1996.

%\bibitem[Greenblatt et al.(1990)]{greenblattetal1990}
%Greenblatt, G. D., Orlando, J. J., Burkholder, J. B. \& Ravishankara, A. R., 1990,
%\jgr, 95, 18577

\bibitem[Hansen \& Sato(2004)]{hansensato2004}
Hansen, J. \& Sato, M., 2004, 
PNAS, 101, 16109.

\bibitem[Hayashida \& Horikawa(2001)]{hayashidahorikawa2001}
Hayashida, S. \& Horikawa, M., 2001,
\grl, 28, 4063

%\bibitem[Hays \& Roble(1968)]{haysroble1968}
%Hays, P. B. \& Roble, R. G., 1968,
%J. Atmos. Sci., 25, 1141

%\bibitem[Hermans et al.(2003)]{hermansetal2003}
%Hermans, C., Vandaele, A. C., Fally, S., Carleer, M., Colin, R., Coquart, B.,
%Jenouvrier, A. \& Merienne, M.-F., 2003, Weakly interacting molecular pairs: 
%Unconventional absorbers of radiation in the atmosphere, C. Camy-Peyret \& A. A. Vigasin,
%Kluwer Academic Publishers: Dordrecht, 193

%\bibitem[Herzberg \& Herzberg(1947)]{herzbergherzberg1947}
%Herzberg, L. \& Herzberg, G., 1947,
%\apj, 105, 353

\bibitem[Hubbard et al.(2001)]{hubbardetal2001}
Hubbard, W. B., Fortney, J. J., Lunine, J. I., Burrows, A., Sudarsky, D. \&
Pinto, P., 2001,
\apj, 560, 413

\bibitem[Hui \& Seager(2002)]{huiseager2002}
Hui, L. \& Seager, S., 2002,
\apj, 572, 540

\bibitem[Kaltenegger \& Traub(2009)]{kalteneggertraub2009}
Kaltenegger, L. \& Traub, W. A., 2009,
\apj, 698, 519

%\bibitem[Keen(1983)]{keen1983}
%Keen, R. A., 1983,
%Science, 222, 1011

\bibitem[L\'eger et al.(2009)]{legeretal2009}
L\'eger, A., Rouan, D., Schneider, J., Barge, P., Fridlund, M., et al., 2009,
 Astron. \& Astrophys., 506, 287L

\bibitem[Link(1959)]{link1959}
Link, F., 1959,
Bull. Astron. Institutes of Czechoslovakia, 10, 105

\bibitem[Link(1962)]{link1962}
Link, F., 1962,
Physics and Astronomy of the Moon,  Ed. by Z. Kopal, pp. 161--229, Academic, New
York

\bibitem[Link(1969)]{link1969}
Link, F., 1969,
Eclipse phenomena in astronomy, pp. 205--225, Springer-Verlag, Berlin

\bibitem[Lissauer et al.(2011)]{lissaueretal2011}
Lissauer, J. J., Fabrycky, D. C., Ford, E. B., Borucki, W. J., Fressin, F., et
al., 2011,
Nature, 470, 53

\bibitem[Manchado et al.(1998)]{manchadoetal1998}
Manchado, A., Fuentes, F. J., Prada, F., Ballesteros, E., Barreto, M., et al.,
1998,
SPIE, 3354, 448

\bibitem[Mat\'e et al.(1999)]{mateetal1999}
Mat\'e, B., Lugez, C., Fraser, G. T. \& Lafferty, W. J., 1999,
\jgr, 104, 30585

\bibitem[Mayor et al.(2009)]{mayoretal2009}
Mayor, M., Bonfils, X., Forveille, T., Delfosse, X., Udry, S., et al., 2009,
Astron. \& Astrophys., 507, 487

\bibitem[Moussaoui et al.(2010)]{moussaouietal2010}
Moussaoui, N., Clemesha, B. R., Holzl\"ohner, R., Simonich, D. M., Bonaccini
Calia, D., et al., 2010,
Astron. \& Astrophys., 511, A31

\bibitem[Pall\'e et al.(2009)]{palleetal2009}
Pall\'e, E., Zapatero Osorio, M. R., Barrena, R., Monta\~n\'es-Rodr\'iguez, P. \&
Mart\'in, E. L., 2009,
Nature, 459, 814

\bibitem[Pall\'e et al.(2011)]{palleetal2011}
Pall\'e, E., Zapatero Osorio, M. R. \& Garc\'ia Mu\~noz, A., 2011,
\apj, 728:19 

\bibitem[Pasachoff et al.(2011)]{pasachoffetal2011}
Pasachoff, J. M., Schneider, G. \& Widemann, T., 2011,
\apj, 141, 112


\bibitem[Plane(2003)]{plane2003}
Plane, J. M. C., 2003,
Chem. Rev., 103, 4963

%\bibitem[Perner \& Platt(1980)]{pernerplatt1980}
%Perner, D. \& Platt, U., 1980,
%\grl, 7, 1053

%\bibitem[Racca(1995)]{racca1995}
%Racca, G. D., 1995,
%Planet. Space Sci., 43, 835

%\bibitem[Rafkin(2012)]{rafkin2012}
%Rafkin, S. C. R., 2012,
%Planet. Space Sci., 60, 147

\bibitem[Rauer et al.(2011)]{raueretal2011}
Rauer, H., Gebauer, S., von Paris, P., Cabrera, J., Godolt, M., et al., 2011,
Astron. \& Astrophys., 529, A8


\bibitem[Rothman et al.(2008)]{rothmanetal2008}
Rothman, L. S., Gordon, I. E., Barbe, A., Benner, D. C., Bernath, P. F.,
Birk, M. et al., 2008,
\jqsrt, 110, 533

%\bibitem[Saari \& Shorthill(1963)]{saarishorthill1963}
%Saari, J. M. \& Shorthill, R. W., 1963,
%\icarus, 2, 115

\bibitem[Seager \& Sasselov(2000)]{seagersasselov2000}
Seager, S. \& Sasselov, D. D., 2000,
\apj, 537, 916

\bibitem[Sidis \& Sari(2010)]{sidissari2010}
Sidis, O. \& Sari, R., 2010,
\apj, 720, 904

\bibitem[Sing et al.(2009)]{singetal2009}
Sing, D. K., D\'esert, J., 
Lecavelier Des Etangs, A., 
Ballester, G. E., Vidal-Madjar, A., et al., 2009,
Astron. \& Astrophys., 505, 891.

\bibitem[Sioris et al.(2010)]{siorisetal2010}
Sioris, C. E., Boone, C. D., Bernath, P. F., Zou, J., McElroy, C. T \& McLinden,
C. A., 2010,
\jgr, 115, D00L14

%\bibitem[Smith \& Newnham(1999)]{smithnewnham1999}
%Smith, K. M. \& Newnham, D. A., 1999,
%Chem. Phys. Lett., 308, 1

\bibitem[Smith \& Newnham(2000)]{smithnewnham2000}
Smith, K. M. \& Newnham, D. A., 2000,
\jgr, 105, 7383

%\bibitem[Smith et al.(2001)]{smithetal2001}
%Smith, K. M., Newnham, D. A. \& Williams, R. G., 2001,
%\jgr, 106, 7541

%\bibitem[Solomon et al.(1998)]{solomonetal1998}
%Solomon, S., Portmann, R. W., Sanders, R. W. \& Daniel, J. S., 1998,
%\jgr, 103, 3847

\bibitem[Swain et al.(2009)]{swainetal2009}
Swain, M. R., Tinetti, G., Vasisht, G., Deroo, P., Griffith, C., et al., 2009,
\apj, 704, 1616

\bibitem[Tanga et al.(2012)]{tangaetal2012}
Tanga, P., Widemann, T., Sicardy, B., Pasachoff, J. M., Arnaud, J., et al., 
Icarus (\textit{in press})

\bibitem[Tinetti et al.(2007)]{tinettietal2007}
Tinetti, G., Vidal-Madjar, A., Liang, M.-C., Beaulieu, J.-P., Yung, Y., et al., 
2007,
Nature, 448, 169

\bibitem[Ugolnikov \& Maslov(2006)]{ugolnikovmaslov2006}
Ugolnikov, O. S. \& Maslov, I. A., 2006,
\jqsrt, 102, 499

\bibitem[Ugolnikov \& Maslov(2008)]{ugolnikovmaslov2008}
Ugolnikov, O. S. \& Maslov, I. A., 2008,
\jqsrt, 109, 378

%\bibitem[van der Werf(2008)]{vanderwerf2008}
%van der Werf, S. Y., 2008,
%Appl. Opt., 47, 153

\bibitem[Vidal-Madjar et al.(2003)]{vidalmadjaretal2003}
Vidal-Madjar, A., Lecavelier des Etangs, A., D\'esert, J.-M., Ballester, G. E.,
Ferlet, R., et al., M., 2003,
Nature, 402, 143

\bibitem[Vidal-Madjar et al.(2004)]{vidalmadjaretal2004}
Vidal-Madjar, A., D\'esert, J.-M., Lecavelier des Etangs, A., 
H\'ebrard, G., Ballester, G. E. et al., 2004, 
\apj, 604, L69

\bibitem[Vidal-Madjar et al.(2010)]{vidalmadjaretal2010}
Vidal-Madjar, A., 
L. Arnold,
D. Ehrenreich,
R. Ferlet,
A. Lecavelier des Etangs et al., 2010,
Astron. \& Astrophys., 523, A57.

%\bibitem[Vigasin \& Slanina(1998)]{vigasinslanina1998}
%Vigasin, A. A. \& Slanina, Z., 1998,
%Molecular complexes in Earth's, planetary, cometary, and 
%interstellar atmospheres, World Scientific: Singapore

\end{thebibliography}
\end{document}